\documentstyle[aps,epsf,amssymb,amsbsy]{revtex}
\def\<{\langle}
\def\>{\rangle}
\newcommand{\tr}{\mathrm{Tr}}

\newcommand{\ud}{\mathrm{d}}
\newcommand{\ue}{\mathrm{e}}
\newcommand{\ui}{\mathrm{i}}
\begin{document}
\draft 
\twocolumn 
\title{Quantum Tomography Via Group Theory}
\author{M. Paini} 
%\address{I.N.F.M., sezione di Pavia, Via Bassi 6, 27100 Pavia, Italy}
\address{Theoretical Quantum Optics Group\\
Dipartimento di Fisica ``A. Volta'' and I.N.F.M., Via Bassi 6, 27100
Pavia, Italy\\paimar@libero.it}
%\date{Received \today}
\maketitle
%\widetext

\begin{abstract}
Amongst the multitude of state reconstruction techniques, the 
so--called
``quantum tomography'' seems to be the most fruitful. In this letter,
I will start by developing the mathematical apparatus of quantum 
tomography and, later, I will explain 
how it can be applied to various 
quantum systems. 
\end{abstract}
\narrowtext
% INTRODUCTION
%%%%%%%%%%%%%%%%%%%%%%%%%%%%%%%%%%%%%%%%%%%%%%%%%%%%%%%%%%%%%%%%%%

Quantum tomography, like all state reconstruction methods, is
concerned with the problem of measuring the density matrix 
$\rho$ of a physical system. The first observations on this topic are
dated 1957 \cite{fano}, but the greatest advancements are a result of
this decade's work. Also the first experiments were performed during this
decade \cite{smithey,kurtsiefer}.
The word tomography stems from medicine, as a consequence of the 
resemblance between the
main formula of the first quantum tomography method, {\it i.e.\ }homodyne
tomography \cite{dariano1,dariano2,dariano3},
and the inverse Radon
transform used in the {\small CAT}
\cite{dariano3,natterer}. Nowadays,
however, 
quantum tomography does not have much in common with its medical ancestor. 
Quantum tomography is a general term, referring to any state measurement
procedure descending from an equation of the same form of equation 
(\ref{teorema}). It 
is a versatile and
mighty technique, as it can be applied to a
great variety of systems and as it includes other methods as special
cases.
\par
Homodyne tomography became a particular case of quantum tomography, when
group representation theory was employed \cite{tesi}. Even though the latter 
led to a breakthrough in quantum tomography,
it was recognized not to be the most general approach. 
There were, in fact, tomographic formulas (formulas like (\ref{teorema})) 
that could not be ascribed
to standard group representation theory. Hence, in order to
include these cases in a general mathematical framework, 
I will introduce some conditions that comprise the definition of
group representation.
Let, then, ${\cal G}$ be a group,
$\cal H$ a separable Hilbert space, 
${\cal B}(\cal H)$ the algebra of linear bounded
operators defined in all $\cal H$, namely Banach algebra \cite{richtmyer},
and $T$ a linear mapping from $\cal G$ to ${\cal B}(\cal H)$ (from now on,
the word linear, when referring to an operator, will always be understood). If we can find 
an irriducible, unitary ray representation $D$ of $\cal G$
(a ray representation is such that
$D(g)D(h)={\ue}^{{\ui}\xi_{gh}}D(g\cdot h)$, with $\xi\in \mathbb R$
\cite{hamermesh}) and six correspondences
$\alpha,\beta:{\cal G}\times {\cal G}\rightarrow {\mathbb R}$ and
$f_{ij}:{\cal G}\rightarrow{\cal G}$, with $i,j=1,2$, satisfying the
equations
\begin{eqnarray}
&&{D}(g)\;{T}(h)={\ue}^{{\ui}\alpha_{gh}}\;{T}(f_{11}(g)
\!\cdot
\!h\!\cdot \!f_{12}(g))\,,
\label{primadef}\\
&&{D}(g)\;{T}^\dag(h)={\ue}^{{\ui}\beta_{gh}}\;
{T}^\dag(f_{21}(g)
\!\cdot
\!h\!\cdot \!f_{22}(g))
\,,\label{secondadef}
\end{eqnarray}
for every $g,h\in {\cal G}$, then we will say that 
$\{T(g)\}$ is a tomographic set.
If $T$ is an irriducible, unitary ray 
representation then $\{T(g)\}$ is a tomographic set (choose
${D}={T}, f_{11}(g)=g,\dots\ $). The converse is not true in general,
so the requirement of $T$ being a representation is more stringent.
When
$\cal H$ is finite--dimensional, 
the hypothesis that $\{T(g)\}$ is a tomographic set
is sufficient to derive (\ref{teorema}),
however the case of ${\rm {dim}}({\cal H})=\infty$ needs a further condition
to make sure that every expression 
converges and can be attributed
a precise mathematical meaning. 
More explicitly, $T$ needs to fulfill the following inequality\cite{nota2} : 
\begin{eqnarray}
\sum_g\;|\<u|{T}(g)|v\>|^2\,<\infty\qquad\forall\, |u\>,|v\>\in 
\cal H.\label{condizione}
\end{eqnarray}
Condition (\ref{condizione}) could be quite a nuisance, since it
must be checked for every couple of vectors $|u\>,|v\>\in 
\cal H$. Fortunately, (\ref{condizione}) is equivalent to
\begin{eqnarray}
\exists \, |u\>,|v\>\in 
{\cal H}\; :\quad\sum_g\;|\<u|{T}(g)|v\>|^2\,<\infty,\label{duesoli}
\end{eqnarray}
as I will demonstrate now. Let us start by admitting the 
validity of (\ref{duesoli}) for two vectors $|w_1\>,|w_2\>$ (eventually
coincident) and
consider the set $V$ of vectors $|v\>\in \cal H$ of the form
\begin{eqnarray}
|v\>=\sum_g\,v_g\,{D}(g)\,|w_i\>,\quad {\mathrm {with }}\;
\sum_g|v_g|<\infty\, ,\label{completezza}
\end{eqnarray}
$i$ being $1$ or $2$. $V$ is a linear manifold 
of $\cal H$ because 
$\alpha|v_1\>+\beta|v_2\>$ belongs to $V$ 
if $|v_1\>,|v_2\>\in V$ and $\alpha,\beta\in \mathbb C$, since
$\sum_g|\alpha\,{v_1}_g\,+\beta\,{v_2}_g\,|\leqslant|\alpha|\sum_g
|{v_1}_g|+|\beta|\sum_g
|{v_2}_g|<\infty$.
The application 
of any ${D}(h)$ to a vector $|v\>\in V$, yields
a vector that still belongs to $V$, in fact ${D}(h)|v\>=
\sum_g\,v_g\,{\ue}^{{\ui}\xi_{hg}}\,{D}(h\cdot g)\,|w_i\>=
\sum_g\,v_{h^{-1}\cdot g}\,{\ue}^{{\ui}\xi_{h(h^{-1}\cdot g)}}
\,{D}(g)\,|w_i\>$, with
$\sum_g|v_{h^{-1}\cdot g}\,{\ue}^{{\ui}\xi_{h(h^{-1}\cdot g)}}
\,|=\sum_g|v_g|<\infty$. The irriducibility of 
the operators ${D}(g)$ implies that $V=\cal H$ or, in other words,
that every vector in $\cal H$ can be written in the form
(\ref{completezza}). 
Hence for every $|u\>,|v\>\in\cal H$, with the help of Cauchy's inequality,
we obtain
\begin{eqnarray}
&&\sum_g\;|\<u|{T}(g)|v\>|^2=\nonumber\\
&&=\sum_g\,\bigg|\,\sum_{g_1,g_2}\,u_{g_1}^*\,v_{g_1}\,\<w_1|
{D}^\dag(g_1)\,{T}^\dag(g)\,{D}(g_2)
|w_2\>\bigg|^2
\leqslant\nonumber\\
&&\leqslant\bigg(\sum_{g_1}|u_{g_1}|\bigg)^{\!\!2}\,
\bigg(\sum_{g_2}|v_{g_2}|\bigg)^{\!\!2}\,\sum_g\;|\<w_1|{T}(g)|w_2\>|^2
<\infty\,,\nonumber
\end{eqnarray}
namely (\ref{duesoli}) $\Rightarrow$ 
(\ref{condizione}) ((\ref{condizione}) $\Rightarrow$ (\ref{duesoli}) is
obvious).
\par
To prove equation (\ref{teorema}) I will work out another identity
first: 
\par\noindent
{\bf {Assertion 1}}
\par\noindent
{\it {If $A$ is a trace--class operator on $\cal H$ 
and $\{T(g)\}$ is a tomographic set and satisfies }}(\ref{duesoli}) 
({\it{or}} (\ref{condizione})) {\it {then
\begin{eqnarray}
{\mathrm {Tr}}\;A
=\frac1{\tilde k}\,\sum_g\,{T}(g)\,A\,{T}^\dag(g),
\label{lemma}
\end{eqnarray}
with ${\tilde k}\equiv\sum_g\;|\<\varphi|{T}(g)|\psi\>|^2$ indipendent
of the choice of the normalized vectors $|\varphi\>,|\psi\>\in \cal H$.}}
\par
\noindent
Proof:
Hypothesis (\ref{condizione}) implies 
\begin{eqnarray}
&&\sum_g\;|\<a|{T}(g)|b\>\,\<c|{T}^\dag(g)|d\,\>|\,\leqslant\nonumber
\\&&\leqslant
\bigg[\sum_g\;|\<a|{T}(g)|b\>|^2\,\sum_g\;|\<d|{T}(g)|c\>|^2
\bigg]^\frac12<\infty\, ,\label{cambiato}
\end{eqnarray}
defining unambiguously 
$\sum_g\;\<a|{T}(g)|b\>\,\<c|{T}^\dag(g)|d\,\>$ 
for all $|a\>$, $|b\>$, $|c\>$, $|d\,\>\in \cal H$, in accordance with
\cite{nota2}.
Since the complete space $\cal H$, as a consequence of Riesz--Fr\'echet 
representation theorem \cite{richtmyer}, is also weakly complete, we
may infer, by virtue of (\ref{cambiato}), that the sequence of the partial
sums $\sum_g^n\,{T}(g)\,|u\>\<v|\,{T}^{\dag}(g)$ is weakly convergent,
as $n\!\to\!\infty$, for all $|u\>,|v\>\in{\cal H}$. The operator
${I}_{uv}\equiv\sum_g\,{T}(g)\,|u\>\<v|\,{T}^{\dag}(g)$ can then be defined
as the weak limit of such sequence, as $n\!\to\!\infty$.
Using equation (\ref{primadef}) and its adjoint and
rearranging the sum we immediately get 
${D}(h)\,{I}_{uv}\,{D}^{\dag}(h)={I}_{uv}$, which,
due to Shur's first lemma (for a proof in the infinite--dimensional
case refer to \cite{tesi}),
is equivalent to ${I}_{uv}=\,k_{uv}\,I$, 
where $k_{uv}\in \mathbb C$ depends on
$|u\>,|v\>$ and $I$ is the identity in $\cal H$. Analogously, equation
(\ref{secondadef}) entails that 
${\widetilde I}_{uv}\equiv\sum_g\,{T}^{\dag}(g)\,|u\>\<v|\,
{T}(g)$ is a multiple of the identity. Given a normalized 
$|w\>\in \cal H$,
$k_{ww}$ may be easily evaluated, in fact $k_{ww}=\<w|k_{ww}\,I|w\>=
\<w|{I}_{ww}|w\>=
\sum_g\;|\<w|{T}(g)|w\>|^2\,,$
where the series could be interchanged with the
inner product because of the way 
$I_{uv}$ is defined.
The constant 
$k_{uv}$ may be expressed in terms of
$k_{ww}$ in the following manner (let $|a\>,|b\>$ be generic vectors):
$\<a|{I}_{uv}|b\>=
\<v|{\widetilde I}_{ba}|u\>
=\<v|u\>\,\<w|{\widetilde I}_{ba}|w\>=
\<v|u\>\,\<a|{I}_{ww}|b\>\, ,$
which means that ${I}_{uv}=\,k_{w,w}\,
\<v|u\>\,I\,$.
The choice $|u\>=|v\>=|\psi\>$ normalized 
and the calculation of the
mean value of the last equation on the normalized vector $|\varphi\>$
produces $\sum_g\;|\<\varphi|{T}(g)|\psi\>|^2=k_{w,w}$, proving
that $\sum_g\;|\<\varphi|{T}(g)|\psi\>|^2$ is indipendent of
the vectors $|\varphi\>,|\psi\>$ (for as long as their norm is 1) 
and will therefore be indicated simply
by $\tilde k$. Schmidt decomposition of a trace--class operator $A$, 
{\it i.e.} 
$A=\sum_ia_i\,|u_i\>\<v_i|,$ where 
$\{|u_i\>\}$ e $\{|v_i\>\}$ are orthonormal sequences and 
$\sum_i a_i<\infty\,$, 
$\,a_i>0\;\,\forall i,$
helps showing that
$\sum_g\,{T}(g)\,A\,{T}^\dag(g)$ is meaningful. It is indeed sufficient 
to check the absolute convergence
of the expression
\begin{eqnarray}
&&\sum_g\,\<a|{T}(g)\,A\,{T}^\dag(g)|b\>=\nonumber\\&&=
\sum_g\sum_ia_i\,\<a|{T}(g)|u_i\>\,\<v_i|{T}^\dag(g)|b\>,
\label{elop}
\end{eqnarray}
for all $|a\>,|b\>\in\cal H$,
to insure the validity of the definition of $\sum_g\,{T}(g)\,A\,{T}^\dag(g)$
as the weak limit of $\sum_g^n\,{T}(g)\,A\,{T}^\dag(g)$, 
as $n\!\to\!\infty$.
The inequality
\begin{eqnarray}
&&\sum_g\;|\<a|{T}(g)|u_i\>\,\<v_i|{T}^\dag(g)|b\>|\,\!\leqslant
\nonumber\\&&\leqslant
\bigg[\sum_g\;|\<a|{T}(g)|u_i\>|^2\,\sum_g\;|\<b|{T}(g)|v_i\>|^2
\bigg]^\frac12\!=\tilde k,\nonumber
\end{eqnarray}
together with $a_i>0$ and $\sum_i a_i<\infty$, guarantees
that the sum of the absolute values of the terms in (\ref{elop}) is
$\qquad\qquad$$\leqslant\tilde k\sum_i\,a_i<\infty$. Because of the absolute 
convergence we can also rearrange the order of the two sums,
obtaining  
the assertion's thesis:
\begin{eqnarray}
&&\frac1{\tilde k}\,\sum_g\,{T}(g)\,A\,{T}^\dag(g)=
\frac1{\tilde k}\sum_i\sum_g\,a_i\,{T}(g)|u_i\>\,\<v_i|\,{T}^\dag(g)=\nonumber
\\
&&=\sum_i\,a_i\,\<v_i|u_i\>\,I={\mathrm {Tr}}\,A\,.\nonumber
\end{eqnarray}

\par 
Now, finally, equation (\ref{teorema}):
\par\noindent
{\bf {Assertion 2}}
\par\noindent
{\it {The operator identity 
\begin{eqnarray}
&&A = \frac{1}{\tilde k}\,\sum_g\,{\mathrm {Tr}}\big[A\,{T}(g)\big]
\, 
{T}^{\dag}(g)\;\label{teorema}
\end{eqnarray}
holds, when $A,T$ and $\tilde k$ are defined as in assertion 1.}}
\par\noindent
Proof:
Let $O$ be an invertible trace--class operator. Using (\ref{lemma}) twice,
it is straightforward
to check that 
\begin{eqnarray}
\sum_g\;\mbox{Tr}[A\,{T}^\dag(g)]\,O\,{T}(g)=
\sum_g\;\mbox{Tr}[\,{T}(g)O]\,{T}^\dagger(g)A
\;.\label{pa1}
\end{eqnarray}
Expanding the trace on the complete orthonormal sequence 
$\{|\varphi_i\rangle\}$,
with the help of equation (\ref{lemma}) again, we may write
\begin{eqnarray}
&&\frac1{\tilde k} \sum_g\;\mbox{Tr}[\,{T}(g)O]\langle \varphi_i|
\,{T}^\dagger(g)A|\varphi_j\rangle=\nonumber\\
&&
=\sum_h\langle \varphi_h|O\,\mbox{Tr}\left[|\varphi_h\rangle\langle
\varphi_i|\right] A|\varphi_j\rangle=\langle \varphi_i|OA|\varphi_j\rangle 
\;\label{pa2}.
\end{eqnarray}
Equations (\ref{pa1}) and (\ref{pa2}) give
$\frac1{\tilde k}\sum_g\mbox{Tr}[A{T}^\dagger(g)]O{T}(g)=OA,$
which is equivalent to (\ref{teorema}) because of the invertibility
of $O$.
\par
Note \cite{aiuto} is devoted to a 
brief comment on some technical features of assertions 1 and 2.

Before we start to examine physical cases, 
I will cast light on some aspects of (\ref{teorema}). It is 
well--known that Hilbert--Schmidt operators form
a Hilbert space, usually denoted as $\sigma_c(\cal H)$, with inner product
$(A,B)_o\equiv{\tr}[A^\dag,B]\;$\cite{kato}, and that the space
$\tau_c(\cal H)$ of trace--class operators is contained in 
$\sigma_c(\cal H)$. If we define 
${P}(g)=
\tilde k^{-\frac12}\,{T}^\dag(g)$, equation 
(\ref{teorema}) can formally be rewritten as
\begin{eqnarray}
A = \sum_g\,(\,{P}(g),A\,)_o\;
{P}(g)\,.\label{completesigma}
\end{eqnarray}
This equality is of simple interpretation: all vectors of
$\tau_c(\cal H)$ can be expanded in terms of a closure relation, resorting
to elements ${P}(g)$ that are not necessarily in $\sigma_c(\cal H)$ but
that belong to a larger set, in the same way that occurs with the expansion
of a vector in terms of generalized  
vectors. 
A comparison will make the situation even clearer. 
If we identify $\sigma_c(\cal H)$ with 
$L^2(\mathbb R)$ (the space of  
square--integrable functions on $\mathbb R$), then $\tau_c(\cal H)$ corresponds
to the space $\cal S(\mathbb R)$ (test functions on $\mathbb R$ 
decreasing rapidly at
infinity) and ${\cal B}({\cal H})$ to the space of tempered
distributions $\cal S'(\mathbb R)$ dual to $\cal S(\mathbb R)$. This
analogy suggests that formula (\ref{completesigma}) is also valid
for $A\in\sigma_c(\cal H)$ if we define the inner product 
$({P}(g),A)_o$ as the limit of $({P}(g),A_n)_o$, with
the sequence of trace--class operators $\{A_n\}$ converging to $A$.
Naturally, not all closure relation arise
from a group.
It is possible to use conditions
similar to (\ref{primadef}) and (\ref{secondadef}) defined on sets that are
not groups and still obtain (\ref{teorema}). Or even more generally, there
are cases of spectral decompositions in $\sigma_c({\cal H})$ 
that do not satisfy anything like
(\ref{primadef}) and (\ref{secondadef}) (for example the eigenvectors of
a self--adjoint operator from $\sigma_c({\cal H})$ to $\sigma_c({\cal H})$
give a closure relation in $\sigma_c({\cal H})$). However, from an 
operative
point of view, these more general approaches to formula (\ref{teorema})
are not very useful. 
Groups 
are indeed simple objects to be dealing with and they produce quite a 
large number
of interesting results.
Moreover, closure relations that do not exhibit a group
structure usually derive from a group. To clarify this point, 
we must
keep in mind that nothing in (\ref{completesigma}) guarantees that 
$\{ {P}(g)\}$ is a complete orthonormal system.
We know
it is complete, because of (\ref{completesigma}), but we cannot be sure
that it is orthonormal with respect to the inner product $(\ ,\ )_o$. 
And, in fact, generally it is not orthonormal, {\it {i.e.}} the 
set $\{ {P}(g)\}$ is often overcomplete. 
If an orthonormal system 
is extracted from the original overcomplete set, the group
properties may disappear
and we would be left with
a spectral decomposition that is not generated by a group, even if it
originated in a group. We will encounter an example of this circumstance
afterwards. 

\par
Equation (\ref{lemma}) is a closure relation itself. Choosing $A=|v\>\<v|$,
with $|v\>$ normalized and arbitrary, it states that $\{P(g)|v\>\}$ is a
complete set in $\cal H$. In other terms, the complete set $\{P(g)|v\>\}$
in $\cal H$ corresponds to the complete set $\{P(g)\}$ in 
$\sigma_c({\cal H})$. Note that this connection is not a consequence
of the specific context 
($\cal G$ being a group and $T$ satisfying (\ref{primadef}) and
(\ref{secondadef})) in which we proved formulas (\ref{lemma}) and
(\ref{teorema}).

\par
So far, I assumed that $\cal G$ was discrete.
Nonetheless, (\ref{lemma}) and (\ref{teorema}) apply to other situations.
It is useful to recall that every unitary irriducible representation of a
compact group is finite--dimensional (meaning that $\cal H$ is
finite--dimensional), whereas every unitary irriducible 
representation of a non--compact group is infinite--dimensional
(with the exception of the trivial representation).
Hence, 
for a finite group or a compact Lie group, the mathematical
problem simplifies: ${\rm {dim}}(\cal H)<\infty$,
$\tau_c({\cal H})=\sigma_c({\cal H})={\cal B}({\cal H})$ and the convergence
of $\tilde k$ is always granted. In particular, for a finite group, 
by tracing both members
of (\ref{lemma}), we immediately recognize that $\tilde k=
\frac{[{\cal G}]}{{\rm {dim}}({\cal H})}$, with $[{\cal G}]$ indicating
the order of $\cal G$. For a compact Lie group, formulas (\ref{lemma})
and (\ref{teorema}) are obtained by substituting $\frac1{[{\cal G}]}
\sum_g$, appearing in (\ref{lemma})
and (\ref{teorema}) for a finite group,  with $\int_{\cal G} {\ud}\mu(g)$, 
where
${\ud}\mu(g)$ is Haar's invariant measure for $\cal G$. Similarly, the formal 
substitution $\sum_g\rightarrow \int_{\cal G} {\ud}\mu(g)$ allows to write
(\ref{lemma}) and (\ref{teorema}) for a non--compact Lie group, with a
discrete group as the starting point. A warning is necessary in this 
case, however.
For non--compact groups $\int_{\cal G} {\ud}\mu(g)$
is not convergent; moreover, 
differently from the compact case, right and
left invariance may correspond to two different ${\ud}\mu(g)$ 
\cite{cornwell}. Only when
they coincide, {\it {i.e.}} only when $\cal G$ is unimodular,
formulas (\ref{lemma}) and (\ref{teorema}) are applicable.

\vskip 1\baselineskip
\par\noindent

Equation (\ref{teorema}) (and (\ref{lemma})) deserves its own
self--existence as a pure
mathematical result, but my initial goal was different.
I was concerned with the physical problem of measuring the density matrix,
and, thus far, 
(\ref{teorema}) does not give us any clue on how to solve such problem. 
I will now show how (\ref{teorema}) is, in actuality, much nearer
to the solution than what may appear. 
If $\cal H$ is
the Hilbert space associated with the physical system under
consideration, then the
density matrix $\rho$ is an element of $\tau_c({\cal H})$ and,
consequently, can be
written in place of
$A$ in formula (\ref{teorema}). Moreover, if each $T(g)$ is self--adjoint 
or is a function of a self--adjoint operator, then we can evaluate the trace 
in (\ref{teorema}) over its
complete set of eigenvectors $\{|g,t\>\}$. This operation yields an
expression containing quantities of the form $\<g,t|\rho|g,t\>$, which can
be interpreted as the probability that  
a measurement
of $T(g)$ gives the eigenvalue $t(g)$ corresponding to $|g,t\>$ (the case
of $t(g)$ degenerate could be treated analogously). These quantities
are, in principle, experimentally accessible and will be
indicated with $p(g,t)$. Formula (\ref{teorema}) then becomes
$\rho=\sum_{g,t}p(g,t)[\frac1{\tilde k}t(g)T^\dag(g)]$, where
if the group is not discrete, or if $T(g)$ has continuous
spectrum, sums must be replaced by integrals (obviously there 
could eventually
be both sums and integrals). The observation that the eigenvectors $|g,t\>$ 
and $|g',t\>$ are not necessarily different even if $g\neq g'$ suggests to
divide $\cal G$ in classes ${\cal G}_i$, requiring the property
that
all $T(g)$ corresponding to the same class have the same eigenvectors.
We would then write
$\rho=\sum_{i,t}p(i,t)\sum_{g\in {\cal G}_i}[\frac1{\tilde k}t(g)T^\dag(g)]
\equiv \sum_{i,t}p(i,t)K(i,t)$, where the operator 
$K(i,t)$ is usually called pattern
function (again, if $i$ is a continuous index then
$\sum_i\rightarrow \int {\ud}\mu(i)$). The formula
\begin{eqnarray}
\rho=\sum_{i,t}\,p(i,t)K(i,t)\label{essenza}
\end{eqnarray}
is the essence of most state reconstruction methods. It
states that measuring the probabilities $p(i,t)$ and calculating
$K(i,t)$ is all that is needed in order to obtain $\rho$. The peculiarity
of quantum tomography is that $K(i,t)$ does not need to be determined by
solving an inverse problem, it is explicitly given by 
$\frac1{\tilde k}\sum_{g\in {\cal G}_i}t(g)T^\dag(g)$.

\vskip 1\baselineskip
\par
It took quite long to work out (\ref{essenza}), but its generality will
show its strength now that we turn to examples.
  
\vskip 1\baselineskip
\par

I will begin with the spin case, as it is the least complex.
The (reduced) spin density matrix 
of one particle with spin $S$ (integer or half--integer)
is defined
in a Hilbert space ${\cal H}_S$, with ${\rm {dim}}({\cal H}_S)
=2S+1$.
The compact group $SU(2)$ is particularly
suited for the case of arbitrary $S$, since there exists an 
irriducible, unitary
representation of $SU(2)$ in every finite--dimensional space.
If $SU(2)$ is parametrized with 
$(\vartheta,\varphi,\psi)$ belonging to
$[0,\pi]\times [0,2\pi)\times [0,2\pi]$
and if $\vec n$ is defined as
$(\cos \varphi\sin
\vartheta,\sin\varphi\sin\vartheta,\cos\vartheta)$, then the operators
$R(\vec n,\psi)={\ue}^{-{\ui}\psi\vec S\cdot\vec n}$, where
$\vec S$ is the spin operator \cite{cornwell}, constitute an irriducible,
unitary representation of $SU(2)$ and, consequently, a tomographic set in
${\cal H}_S$. Formula (\ref{essenza}) then becomes
\begin{eqnarray}
\rho=
\int_{\mathit \Sigma}{{\ud}\Omega}_{\vec n}\sum_{m=-S}^{S}p(\vec n,m)
\;K_S(\vec n,m)\,,\label{classica}
\end{eqnarray}
where ${{\ud}\Omega}_{\vec n}$ is the area element of the unitary
spherical surface ${\mathit \Sigma}$,
$K_S(\vec n,m)$ is the pattern
function given by 
$(2S+1)\int_0^{2\pi}{\ud}\psi\,
\frac{\sin^2\frac\psi2}{4\pi^2}\,
{\ue}^{{\ui}\psi(\vec S\cdot\vec n-m)}$, and
$p(\vec n,m)$ is the probability that $m$ is the result of a 
measurement of $\vec S\cdot\vec n$. 
The calculation of $K_S(\vec n,m)$,
the experimental apparatus needed to measure $p(\vec n,m)$ and some
numerical simulations can be found in \cite{tesi,darianomacconepaini}.
Because ${\rm{dim}}({\cal H}_S)<\infty$, it is evident that the
tomographic set $\{R(\vec n,\psi)\}$ is overcomplete.
It is then possible to choose a finite number of operators $R(\vec n,\psi)$ 
and still obtain a closure relation in $\sigma_c({\cal H}_S)$. 
As previously mentioned,
the operation of extraction of a smaller complete set from
the entire set $\{R(\vec n,\psi)\}$ usually 
produces a class of operators 
not corresponding to a group. However, this need not always be the case.
The dihedral and tetrahedral subgroups of $SU(2)$ \cite{hamermesh} 
can be represented with
a finite number of operators $R(\vec n,\psi)$, producing
tomographic formulas for the cases of $S=\frac12$ and $S=1$
(for explicit formulas and some
further considerations refer to \cite{tesi,darianomacconepaini}).
Unfortunately, there is only a finite number of finite subgroups of $SU(2)$;
furthermore,
the tomographic set associated with the tetrahedral group 
is still overcomplete. Therefore, waiving the group structure is a necessity
for the obtainment of a complete orthonormal system in 
$\sigma_c({\cal H}_S)$ for a generic S.
This does not mean that we have to completely 
give up the use of the operators
$R(\vec n,\psi)$.
Knowing in advance that 
${\rm {dim}}(\sigma_c({\cal H}))=(2S+1)^2$,
we can just choose $(2S+1)^2$ linearly independent operators 
$R(\vec n,\psi)$, that do not need to form or correspond to a group,
and apply Gram--Schmidt orthonormalization procedure:
$B_1\!\equiv\!\frac{R(\vec n_1,\psi_1)}{\|R(\vec n_1,\psi_1)\|_o}$,
$B_2\!\equiv\!\frac{R(\vec n_2,\psi_2)-(B_1,R(\vec n_2,\psi_2))_o\;B_1}
{\|R(\vec n_2,\psi_2)-(B_1,R(\vec n_2,\psi_2))_o\;B_1\|_o}$,$\ \dots$,
with $\|O\|_o\!\equiv\!\sqrt{(O,O)_o}\,,\;\forall \,O\!\in\!
\sigma_c({\cal H})$.
By definition, $\{B_i\},i\!=\!1,2,\dots,(2S+1)^2,$ is a basis in
$\sigma_c({\cal H})$. Nevertheless also $\{B_i^\dag\}$ is a basis
in $\sigma_c({\cal H})$, 
therefore every $A$ in $\sigma_c({\cal H})$
can be decomposed as $A\!=\!\sum_i(B_i^\dag,A)_o\;B_i^\dag
\!=\!\sum_i{\tr}[B_i\,A]\,B_i^\dag\,$.
Because the operators $B_i$ are linear combinations of the operators
$R(\vec n_i,\psi_i)$, with a little algebra we get
$A\!=\!\sum_i{\tr}\big[R(\vec n_i,\psi_i)\,A\big]\,{\cal R}_i\,$,
where the $(2S+1)^2$ operators ${\cal R}_i$ are linear combinations of the
operators $R^\dag(\vec n_i,\psi_i)$ as a result of the reorganization of
the sum on $i$. One may check that $({\cal R}^\dag_i,R(\vec n_j,\psi_j))_o=
\delta_{ij}$ (Kronecker's delta), which means that $\{{\cal R}_i\}$ is the
dual basis of $\{R(\vec n_i,\psi_i)\}$ in $\sigma_c({\cal H})$. 
Calculating the last
trace on the eigenstates $|\vec n_i,m\>$, associated with
the eigenvalue ${\ue}^{-{\ui}\psi_i\,m}$ of the operators 
$R(\vec n_i,\psi_i)$, with $A=\rho$, we attain a finite version of
(\ref{classica}):
\begin{eqnarray}
\rho=
\sum_{i=1}^{(2S+1)^2}\!\sum_{m=-S}^{S}p(\vec n_i,m)
\;{\cal K}_S(i,m)\,,\label{classicafinita}
\end{eqnarray}
with ${\cal K}_S(i,m)={\ue}^{-{\ui}\psi_i\,m}\;{\cal R}_i$ (the
suffix $S$ in ${\cal K}_S(i,m)$ is a remainder of the dependence of the
operators ${\cal R}_i$ on the dimension of ${\cal H}_S$).
Equations having the same form of (\ref{classica}) or (\ref{classicafinita})
are quite common in the problem of spin state reconstruction
%Several spin state 
%reconstruction techniques are based on equations of the same form as
%(\ref{classica}) or (\ref{classicafinita})
(for example
\cite{dodonov,safonov,newton}).
Incidentally,
we might also observe that the orthonormalization of $(2S+1)^2$
linearly independent 
projectors
$|\vec n_i,S\>\<\vec n_i,S|$, $i=1,2,\dots,(2S+1)^2$, instead of the
operators $R(\vec n_i,\psi_i)$, leads to the same results obtained
by Amiet and Weigert \cite{weigert}. 

\par
The Hilbert space ${\cal H}_{C}$ 
associated with a system of $n$ spins is given
by the tensor product of the $n$ single--spin spaces. If we were to
write (\ref{teorema}) for the elements of $\sigma_c({\cal H}_{C})$, we would
tempted to choose ${\cal G}=SU(2)\times\dots
\times SU(2)$ ($n$ times) and $D(g)=T(g)=\bigotimes_{i=1}^n
R(\vec n_i,\psi_i)$, with $g\in {\cal G}$ ($\times$ and
$\bigotimes$ denote respectively the direct product of groups and the 
tensor product of operators). This choice would actually give a 
valid closure relation in $\sigma_c({\cal H}_{C})$ (equation
(\ref{teorema})), 
but the corresponding
probabilities in equation (\ref{essenza}) would require measurements
on single components of the system, which are not always possible. This
difficulty, at least for systems of spins $\frac 12$, can be overcome with
a different approach, as illustrated in \cite{paini2}.

\vskip 1\baselineskip
\par\noindent
 
As a model for systems associated with infinite--dimensional Hilbert spaces,
we can take the space ${\cal H}_O$ of one mode of the 
electromagnetic field. Although the problem is mathematically identical for
other systems (for example, ${\cal H}_O$ is isomorphic to the space of
a spinless, non--relativistic particle in one dimension), 
quantum optics gives the unique
possibility of measuring the equivalent of linear combinations of
position and momentum, namely the so--called quadratures 
$X_\phi\equiv\frac 12(a^\dag
{\ue}^{{\ui}\phi} + a\,{\ue}^{-{\ui}\phi})$, with $\phi\in {\mathbb R}$ and 
$a$ and $a^\dag$ indicating
the annihilation and creation operators respectively \cite{darianoancora}.
This opportunity can be exploited by choosing the non--compact 
Lie group of translations
in the complex plane, with elements $\alpha\in \mathbb C$, as $\cal G$.
Since the 
displacement operators ${\mathsf D}(\alpha)=\exp(\alpha\,a^\dag-
\alpha^*\,a)$ form an irriducible, unitary ray representation of $\cal G$,
such that $\int_{\mathbb C} {\ud}^2\alpha\;|\<0|{\mathsf D}(\alpha)|0\>|^2=\pi$
(${\ud}^2\alpha\equiv{\ud ({\rm {Re}}\,\alpha)}\,{\ud ({\rm {Im}}\,\alpha)}$ 
is the invariant measure of
the unimodular
group of translations in the plane \cite{hamermesh} and $|0\>$ is the
vacuum state),
we can set $D(\alpha)=T(\alpha)={\mathsf D}(\alpha)$. For
the purpose
of writing (\ref{essenza}), we should, however, express the
tomographic set
in terms of quadratures. This can be achieved by 
parameterizing $\cal G$ with
$k\in \mathbb R$ and $\phi\in [0,\pi)$, related to $\alpha$ by 
the equation $\alpha=\frac {\ui}2 k\,{\ue}^{{\ui}\phi}$, 
since
$T(\phi,k)\equiv T(\alpha(\phi,k))=
{\ue}^{{\ui}kX_\phi}$. 
Equation (\ref{essenza}) for this case then reads
\begin{eqnarray}
\rho=\int_0^\pi {\ud}\phi\int_{-\infty}^{+\infty} {\ud}x\,p(\phi,x)\,
{\mathsf K}(\phi,x)\,,\label{ottica}
\end{eqnarray}
where $p(\phi,x)$ is the probability that measuring $X_\phi$ we get $x$ and
${\mathsf K}(\phi,x)\!=\!
\frac1\pi\int_{-\infty}^{+\infty} {\ud}k 
\,\frac{|k|}4\,{\ue}^{{\ui}k (x-X_\phi)}$ (note that $\tilde k=\pi$). 
Equation (\ref{ottica}) is the fundamental formula of homodyne tomography,
which I will not be discussing here, because the literature on it is already
abundant
(\cite{dariano1,dariano2,dariano3} and references therein).
\par
Homodyne tomography is not the only technique that allows to reconstruct
the state $\rho$ of one mode of the electromagnetic field. K. Banaszek and
K. W\'odkiewicz showed how $\rho$ could be determined by measuring, for 
every $\alpha\in\mathbb C$ and $n\in\mathbb N$,
the probability $p(\alpha,n)$ 
that $n$ is the number of photons in the state
${\mathsf D}(\alpha)\rho\,{\mathsf D}^\dag(\alpha)$ \cite{banaszek}.
The same result
can be recovered with equation (\ref{essenza}), if we maintain the choices 
$\cal G=\;$\{Translations in the complex plane\} and $D(\alpha)=
{\mathsf D}(\alpha)$, but we select the tomographic set 
$\{{\mathsf D}^\dag(\alpha)\,{\ue}^{{\ui}y
a^\dag a}{\mathsf D}(\alpha)\}$, where $y$ can be any real number that
is not a multiple of $2\pi$. It is not hard to check that
(\ref{primadef}) and (\ref{secondadef}) are satisfied \cite{unultimanota}
and that $\int_{\mathbb C} {\ud}^2 \alpha |\<0|
{\mathsf D}^\dag(\alpha)\,{\ue}^{{\ui}y
a^\dag a}{\mathsf D}(\alpha)|0\>|^2=\frac{\pi}{2(1-\cos y)}$. Then
equation (\ref{essenza}) is simply
\begin{eqnarray}
\rho=\int_{\mathbb C} {\ud}^2\alpha\;\sum_{n=0}^{+\infty}\;p(\alpha,n)\;
{\mathrm K}_y(\alpha,n)\,,\label{bana}
\end{eqnarray} 
with ${\mathrm K}_y(\alpha,n)=\frac{2(1-\cos y)}{\pi}\,{\mathsf D}^\dag
(\alpha)\,{\ue}^{{\ui}y(n-
a^\dag a)}{\mathsf D}(\alpha)$.

\par
Differently from (\ref{classica}), neither (\ref{ottica}) nor (\ref{bana})
can be simplified by extracting a complete subset from the entire
tomographic set,
since both $P_1(\alpha)\equiv
\pi^{-\frac12} {\mathsf D}^\dag(\alpha)$ and
$P_2(\alpha)\equiv(\frac{\pi}{2(1-\cos y)})^{-\frac12}
{\mathsf D}^\dag
(\alpha)\,{\ue}^{-{\ui}y
a^\dag a}{\mathsf D}(\alpha)$ form orthonormal systems:
$(P_1(\alpha),P_1(\alpha'))_o=(P_2(\alpha),P_2(\alpha'))_o=
\delta({\mathrm {Re}}\,\alpha
- {\mathrm {Re}}\,\alpha')\;\delta({\mathrm {Im}}\,\alpha
- {\mathrm {Im}}\,\alpha')$, with $\delta$
indicating Dirac's delta.

\vskip 1\baselineskip
\par
 Some examples have shown how different state reconstruction problems can
be treated as particular cases of a general method, which can
be summarized in (\ref{teorema}) and in the consequent (\ref{essenza}).
Although the
theory developed in this letter is quiete comprehensive, there is still
space for further generalizations. One possibility is to assume that the
mappings $D$ and $T$ are defined on two distinct groups,
changing (\ref{primadef}) and (\ref{secondadef})
appropriately. Apparently, this generalization would produce 
%This seems not to
%be a sterile generalization, since, apparently,
other interesting
tomographic formulas. For example, the state $\rho$ of one optical mode
could be determined by measuring only
the presence (or the absence) of photons in the state
${\mathsf D}(\alpha)\rho\,{\mathsf D}^\dag(\alpha)$, for every
$\alpha\in {\mathbb C}$.
%would be generated.

\vskip 1\baselineskip
\par
My gratitude goes to Prof. G. M. D'Ariano and L. Maccone of the
Quantum Optics Group of Pavia.  This work,
without their collaboration, would not have
existed.

%This work is dedicated to Matteo, a friend who has passed away too soon.


\begin{thebibliography}{99}
\bibitem{nota2} In what follows I assume that $\cal G$ is discrete. This
is really not a limitation, and I will discuss later how
all results remain valid for other types of group. Also, note that 
the sum on all elements $g\in\cal G$ in (\ref{condizione})
is well--defined. In fact, since $\cal G$ is discrete, there exists a 
$1:1$ mapping $F$ between $\cal G$ and $\mathbb N$,
therefore the sum over $\cal G$ can be written as an ordinary numerical series.
The assumption (\ref{condizione}), namely the absolute convergence of
the series in (\ref{condizione}), guarantees that the value of the
series does not depend on the choice of the bijective correspondence 
$F$.
\bibitem{unultimanota} The set $\{D^\dag(\alpha)\,{\ue}^{{\ui}y
a^\dag a}\,D(\alpha)\}$ is a particular case of a class of 
tomographic sets. In fact, it would be possible to show that,
given an 
irriducible, unitary ray representation $D$ of $\cal G$, the operators
$D^\dag(g)\,T_0\,D(g)$ form a tomographic set, whenever $\cal G$ is Abelian
and $T_0$ is such that $T_0\,D(g)
={\ue}^{{\ui} \eta_g} D({\tilde f}(g))\,T_0$,
for all $g\in \cal G$, with $\eta:\cal G\rightarrow \mathbb R$ and with 
$\tilde f:\cal G\rightarrow \cal G$ satisfying certain rather involved
conditions that will
not be investigated here.
\end{thebibliography}
\end{document}